\documentclass{aa}  
\usepackage{subfigure}
\usepackage{threeparttable}
\usepackage{parskip}
\usepackage{float}
\usepackage{hyperref}
\usepackage{graphicx}
\usepackage{color}
\usepackage{url}
\usepackage[normalem]{ulem}
\usepackage{txfonts}
\usepackage{changes} %
\definechangesauthor[name={Lu}, color=red]{lu} %

\newcommand{\nicer}{{NICER\/}} %
\newcommand{\cxo}{{\it Chandra\/}}

\newcommand{\maxi}{{MAXI\/}} %

\def\be{\begin{equation}} 
\def\ee{\end{equation}}

\begin{document}

   \title{A transition from mixed-fuel to pure-helium thermonuclear bursts in Terzan 5 X--3/Swift J174805.3--244637}
      \titlerunning{Thermonuclear X-ray bursts of Terzan 5 X--3}
   \authorrunning{Zhang et al.}

\author{Lei Zhang\inst{1}
        \and
        Zhaosheng Li\inst{1}
        \thanks{Corresponding author}
       \and
    Yuanyue Pan\inst{1}
    \thanks{Corresponding author}
       \and 
       Wenhui Yu\inst{1}
             \and
    Yupeng Chen\inst{2}
    \and
    Yue Huang\inst{2}
    \and
    Mingyu Ge\inst{2}
    \and
    Shu Zhang\inst{2}
          }
   \offprints{Z. Li}

   \institute{Key Laboratory of Stars and Interstellar Medium, Xiangtan University, Xiangtan 411105, Hunan, P.R. China\\ \email{lizhaosheng@xtu.edu.cn, panyy@xtu.edu.cn}
               \and
Key Laboratory of Particle Astrophysics, Institute of High Energy Physics, Chinese Academy of Sciences, 19B Yuquan Road, Beijing 100049, China              }
   \date{Received XX; accepted XX}

\abstract{We presented a detailed analysis of seven thermonuclear X-ray bursts from Terzan~5~X-3/Swift J174805.3--244637, detected by \nicer\ during the source's 2023 outburst. Our analysis reveals a clear evolution of burst properties, identifying four non-photospheric radius expansion (non-PRE) bursts, one PRE candidate occurring in a mixed hydrogen/helium environment, and two powerful PRE bursts from pure helium ignition. 
The time-resolved burst spectra were well described by a model including a variable persistent emission component, quantified by a factor $f_a$, due to the Poynting-Robertson drag.  The strength of this interaction scales with burst luminosity: the enhancement is absent ($f_a \approx 1$) in the faintest bursts, becomes modest ($f_a \approx 1.5-2$) for the more luminous non-PRE burst and the PRE candidate, and is very strong ($f_a \approx 6-8$) during the pure-helium PRE bursts.
This observed transition from mixed-fuel to pure-helium burning as the local mass accretion rate dropped below $\sim$10\% of the Eddington limit, $\dot{m}_{\rm Edd}$, aligns with theoretical predictions. We verified this scenario with two independent methods. First, at the known distance to Terzan 5, the touchdown luminosities of both the pure helium PRE bursts and the mixed-fuel PRE candidate are consistent with reaching their respective, composition-dependent Eddington limits on the same plausible, massive neutron star of $\sim$2\,$M_\odot$.  Second, the observed recurrence times of the non-PRE bursts were consistent with predictions for mixed-fuel burning. }
   \keywords{X-ray bursts; X-ray sources; Neutron stars; Low-mass X-ray binary stars           }

   \maketitle

\section{Introduction}

\label{Sec:intro}
In low-mass X-ray binaries (LMXBs), matter from a low mass companion star transfer onto a compact object via disk accretion \citep{Frank92}. In systems hosting a neutron star (NS), the accumulation of accreted material on the stellar surface can lead to unstable thermonuclear burning, observed as type I X-ray bursts \citep[see e.g.,][for reviews]{Lewin93,Strohmayer06,Galloway21}. For hydrogen-rich accretion, the burst properties are critically dependent on the fuel composition, which is primarily governed by the local mass accretion rate, $\dot{m}$ (in Eddington units, $\dot{m}_\mathrm{Edd}$). At very low rates ($\dot{m} \lesssim 0.01\,\dot{m}_\mathrm{Edd}$), stable hydrogen burning is inefficient. At moderate rates ($0.01 \lesssim \dot{m} \lesssim 0.1\,\dot{m}_\mathrm{Edd}$), hydrogen burns stably to helium, forming a helium layer that later ignites pure helium burning via the 3$\alpha$ process. At higher rates ($0.1 \lesssim \dot{m} \lesssim 1\,\dot{m}_\mathrm{Edd}$), ignition occurs before hydrogen is depleted, resulting in bursts from a mixed-fuel
environment. Finally, at near-Eddington rates, burning can become (marginally) stable, the burst activities are quenched \citep{Fujimoto81,Strohmayer06,Li21}.

The time-resolved spectra of type I X-ray burst can be well described by a blackbody model. 
However, significant deviations from a simple blackbody often occur. More complex models are frequently required to account for continuum variations, such as disk reflection or a temporary enhancement of the persistent emission due to the Poynting-Robertson drag \citep{Walker92, Worpel13, Degenaar18,Zhao22,2023A&A...670A..87L,2024A&A...683A..93Y,2025A&A...696A.139Y}. Furthermore, spectral features have been observed, including absorption and emission lines, as well as absorption edges \citep{Li18, Strohmayer19, Lu24, 2025ApJ...982...18P,2025A&A...696A.139Y}. In some X-ray bursts where the peak luminosity reaches the Eddington luminosity, the radiation pressure on the accreted material at the NS surface exceeds the gravitational potential, driving photospheric radius expansion \citep[PRE;][]{Lewin93}. They were used as standard candles for distance determination \citep{Kuulkers03}, and analysis of their touchdown moment and subsequent cooling tail provides a method for constraining the NS masses and radii \citep[e.g.,][]{Sztajno87,Ozel09,Suleimanov17,Poutanen14,Li18}.

Terzan 5 is a dense and massive globular cluster located near the Galactic Center, with a
precisely measured distance of $6.62 \pm 0.15$ kpc \citep{2022ApJ...941...22M}. The cluster is rich in X-ray sources, including a large population of quiescent NS LMXBs \citep{2006ApJ...651.1098H, 2015MNRAS.451.2071D} and two other
well-known transients, EXO 1745--248 and the accreting millisecond pulsar
IGR J17480--2446 \citep{2002ApJ...577..337S, 2011ApJ...731L...7M}. The target of this work, Terzan 5 X--3 (also known as Swift J174805.3--244637), was discovered as a new transient during an outburst in July 2012. Follow-up observations with \cxo\ confirmed it as the third transient NS LMXB in the cluster \citep{2012ATel.4242....1W, 2012ATel.4249....1H, 2012ATel.4264....1A}. During that outburst, a single type I X-ray burst with a long decay timescale ($\tau \approx 16-29$\,s) was detected, which established that the system hosts a
hydrogen-rich companion star \citep{2014ApJ...780..127B}. A new outburst from the source was detected by \maxi/GSC on 2023 February 27, with its association with Terzan 5 X--3 quickly confirmed by \textit{Swift}/XRT and \cxo\ \citep{2023ATel15917....1N, 2023ATel15919....1K, 2023ATel15953....1H}. Throughout this most recent outburst, multiple Type~I X-ray bursts were observed by both \nicer\ and \cxo\  \citep{2023ATel15922....1S, 2023ATel15953....1H}.

In this work, we present a detailed analysis of seven X-ray bursts from Terzan 5 X--3 observed by \nicer\ during its 2023 outburst. The observations and outburst
properties are described in Section~\ref{Sec:observe}. Our analysis of the persistent
and time-resolved burst spectra is detailed in Section~\ref{sec:spec_analysis}. We
discuss the implications for the burst fuel composition in Section~\ref{Sec:discussion}
and provide a summary of our findings in Section~\ref{Sec:conclusion}.

\section{Observations}
\label{Sec:observe}

The 2023 outburst from Terzan 5~X--3 was first detected by \maxi\ at MJD 60002, while \nicer\ carried out observations starting at MJD 60004. We analyzed the \nicer\ data for this source, which include Obs. IDs 6203670101--6203670124, i.e., 24 observations with total unfilter exposure time of 131.5 ks.

We note that \nicer\ is a nonimaging instrument with a relatively large
field of view of $(30')^2$, which encompasses Terzan~5. This region is known to host numerous other X-ray sources, including several NS LMXBs. However, based on the X-ray monitoring of the cluster by \maxi, all other known transients were in a quiescent state during the 2023 outburst \citep{2023ATel15917....1N, 2023ATel15919....1K, 2023ATel15953....1H}. We can therefore confidently attribute the observed flux and bursting activity to Terzan 5 X--3, and any contribution from contaminating sources is considered negligible.

We processed \nicer\ data using \texttt{HEASOFT v6.34} and the NICER Data Analysis Software (\texttt{NICERDAS}).
We first extracted the 1-s binned light curves from calibrated unfiltered (UFA) event files to avoid missing type I X-ray bursts \citep[see e.g.,][]{2023A&A...670A..87L, 2024A&A...683A..93Y}. Seven type I X-ray bursts were identified based on the shape of light curves. Then, we used the \texttt{nicer-l2} tool to extract cleaned event files by applying the standard filtering criteria. The 1-s binned light curves in 0.5--10 keV were also extracted from the cleaned files using \texttt{nicerl3-lc}. A comparison of the UFA and cleaned data confirmed the necessity of our initial approach; for example, the rising phase of burst~\#2 and the later decay phase of burst \#4 were present in the UFA data but absent from the standard cleaned event files.

We also extracted light curves with 64 s time bins in three energy bands, 0.5--10\,keV, 2.0--3.8\,keV, and 3.8--6.8\,keV,  using the \texttt{nicerl3-lc} command and calculated the hardness ratio between the 2.0--3.8\,keV and 3.8--6.8\,keV bands.

\subsection{The 2023 outburst profile}
\label{sec:outburst}
\begin{figure}
    \sidecaption
        \includegraphics[width=9cm]{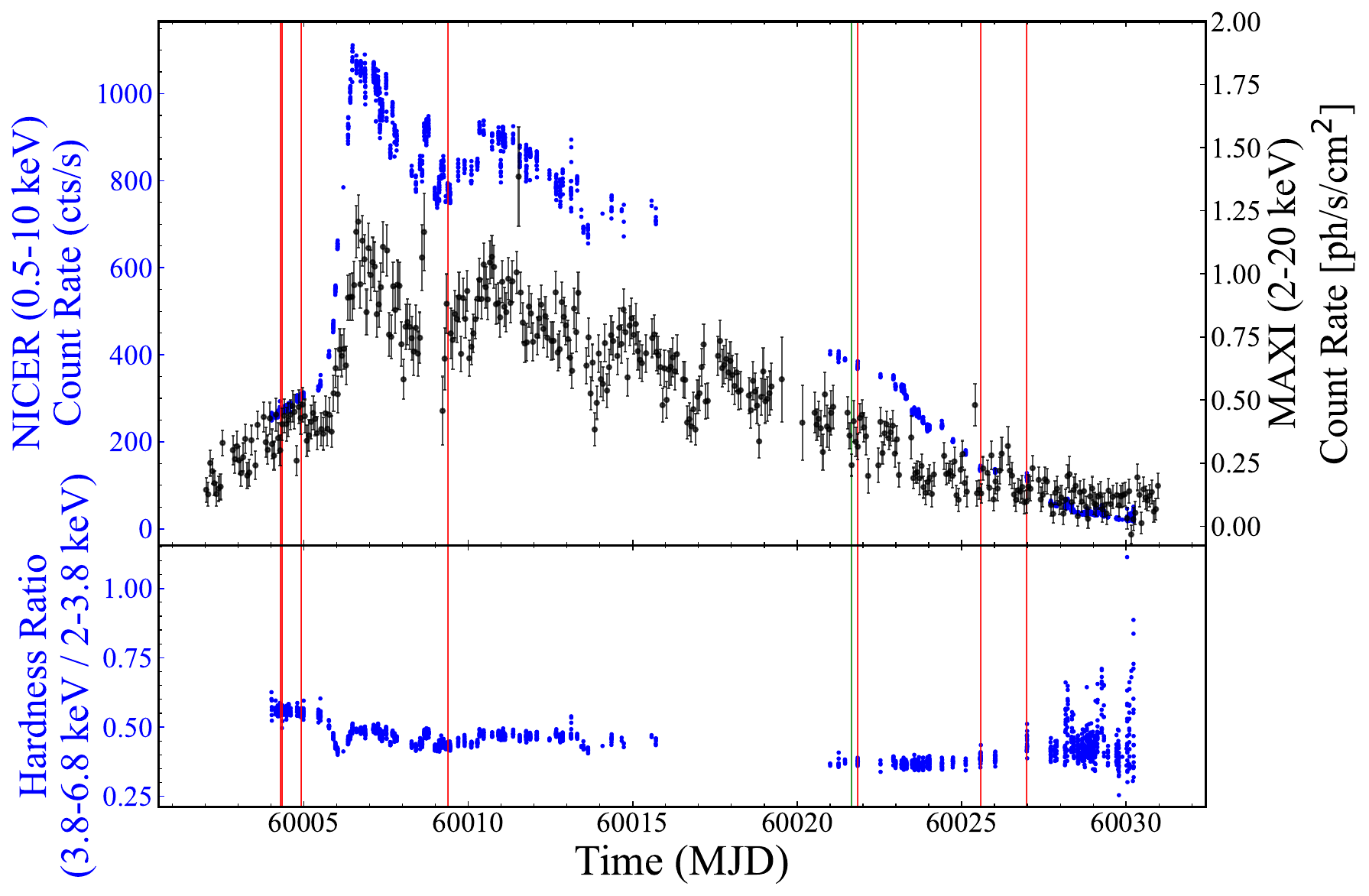} 
\caption{    
The 2023 outburst began on MJD 60002 (2023 February 27). The upper panel shows the light curves from \nicer\ observations (0.5--10 keV, blue points), and  \maxi\ (2--20 keV, black points). The lower panel shows the hardness ratio, defined as the count rate ratio between 3.8–6.8 keV and 2.0–3.8 keV. Each data point represents 64 seconds of \nicer\ data, with all bursts excluded. The red dashed lines indicate the type I X-ray bursts observed by \nicer\, while the green dashed line represents the burst observed by \cxo\ \citep{2023ATel15953....1H}.}
\label{Fig:lc}
\end{figure}

In Fig.~\ref{Fig:lc}, we presented the light curve and hardness ratio evolution with 64 s time bins during the 2023 outburst. In addition, we also downloaded the \maxi\ data from the official webpage \footnote{\url{http://maxi.riken.jp/star_data/J1748-248/J1748-248.html}}. The 2--20 keV outburst light curve binned with one orbit are shown as black points in Fig.~\ref{Fig:lc}. Type~I X-ray bursts from \nicer\ were excluded from the light curve. The onset time of burst detected by \nicer\ and \cxo\ are marked as red and greed dashed lines, respectively \citep{2023ATel15953....1H}. 

The light curves from \nicer\ and \maxi\ are broadly consistent, with no
evidence for dipping or eclipsing activity. The overall outburst profile, best traced by the continuous monitoring from \maxi, exhibits a multi-stage evolution. The outburst began with a slow-rise phase lasting between $\sim$MJD~60001 to 60004, during which the source brightened steadily while in the hard spectral state. This was followed by a fast-rise phase of $\sim$1.5 days, where the count rate increased rapidly. The outburst peaked around MJD~60006. The subsequent decay phase, spanning over $\sim24$ days, can be characterized as a gradual decline with fluctuations. This complex decay behavior is also 
shown in the  \nicer\ observations. Finally, the count rate declined towards its quiescent level.

\begin{figure}
\sidecaption
    \includegraphics[width=9cm]{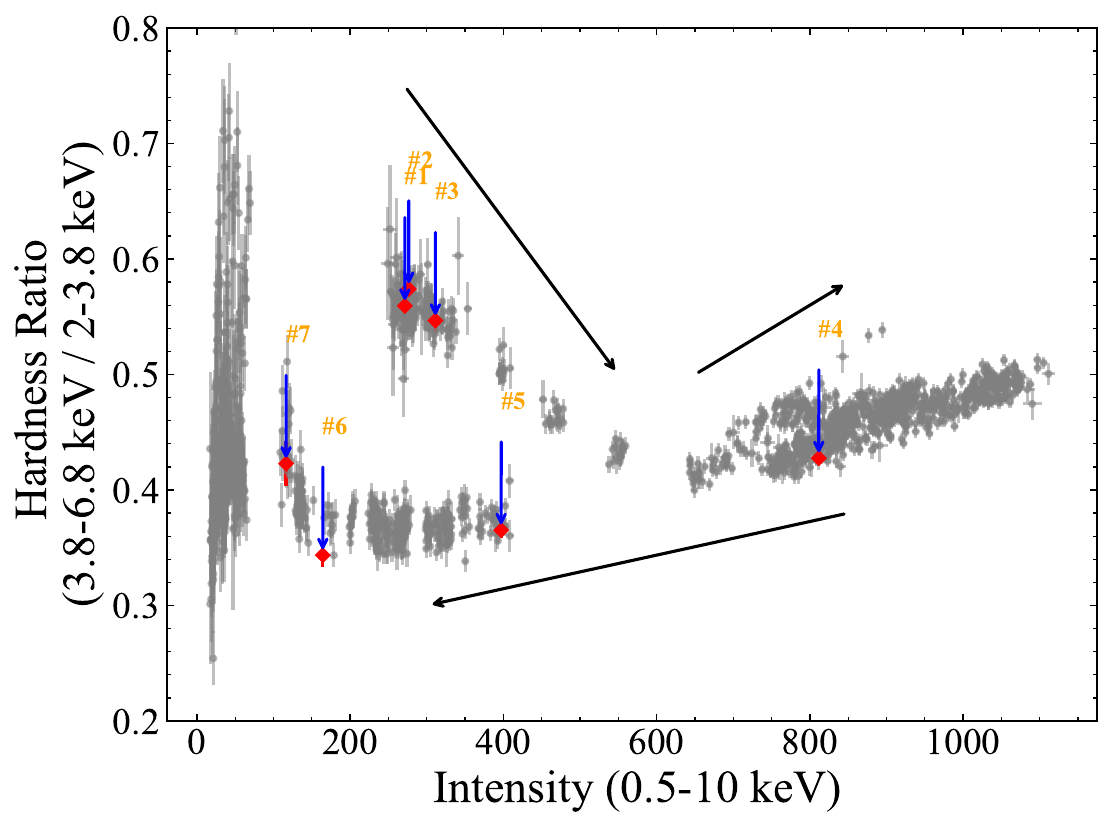}
\caption{
The hardness-intensity diagram (HID) of the Terzan 5 X--3 during the 2023 outburst from \nicer\ observations.  All bursts are removed, and each point represents a segment of 64 s. The HID of the persistent emission before each \nicer\ X-ray burst are marked as rad diamond points. The black arrow indicates the direction of HID evolution. }
\label{Fig:HID}%
\end{figure}

The hardness-intensity diagram (HID) of the 2023 outburst reveals the characteristic clockwise, a ``q''-shaped pattern typical of transient LMXBs, see Fig.~\ref{Fig:HID}. The outburst began in a spectrally hard state (hardness ratio  $\gtrsim0.6$), moving downwards to the right in the HID as it brightened. At a peak intensity of over 1000 cts/s, the source transitioned to a soft state, marked by a rapid drop in hardness to a ratio of $\sim0.4$. It remained on this soft-state for the majority of the outburst decay, with the intensity decreasing while the hardness remained low. Finally, at low intensity, the source transitioned back to the hard state.  The state of the persistent emission immediately prior to each burst is marked by a red diamond. The first three bursts were triggered during the hard state, while the subsequent four occurred during the soft state.  We note that no bursts were detected at the highest source intensities, above $\sim800$ cts/s. This absence could be due to observational gaps in the data coverage, or it could indicate that at the highest accretion rates, the thermonuclear burning on the NS surface became stable, thus suppressing the production of type I X-ray bursts \citep[see e.g.,][for the case of 4U 1820--30]{2024A&A...683A..93Y}.

\subsection{The X-ray bursts Light Curves}
\label{Sec:burst_lc}

To analyze the temporal properties of the seven bursts, we extracted their light curves with a 0.5 s time bins in the 0.5--10\,keV. For each burst, the underlying persistent emission was estimated from a 100-s pre-burst interval and subtracted to obtain the net burst light curve. The one exception was burst~\#2, which was identified in the UFA file and lacked pre-burst data; for this event, a 100-s post-burst interval was used for the background estimate. We defined the burst onset and end times as the epoch where the net count rate rose above and returned to 1.5 times the pre-burst average level, respectively. The rise time, $\Delta t_\mathrm{rise}$, was measured as the duration from onset to the time bin that first reached 90\% of the peak count rate. 

\begin{figure*}
\centering
\includegraphics[width=18cm]{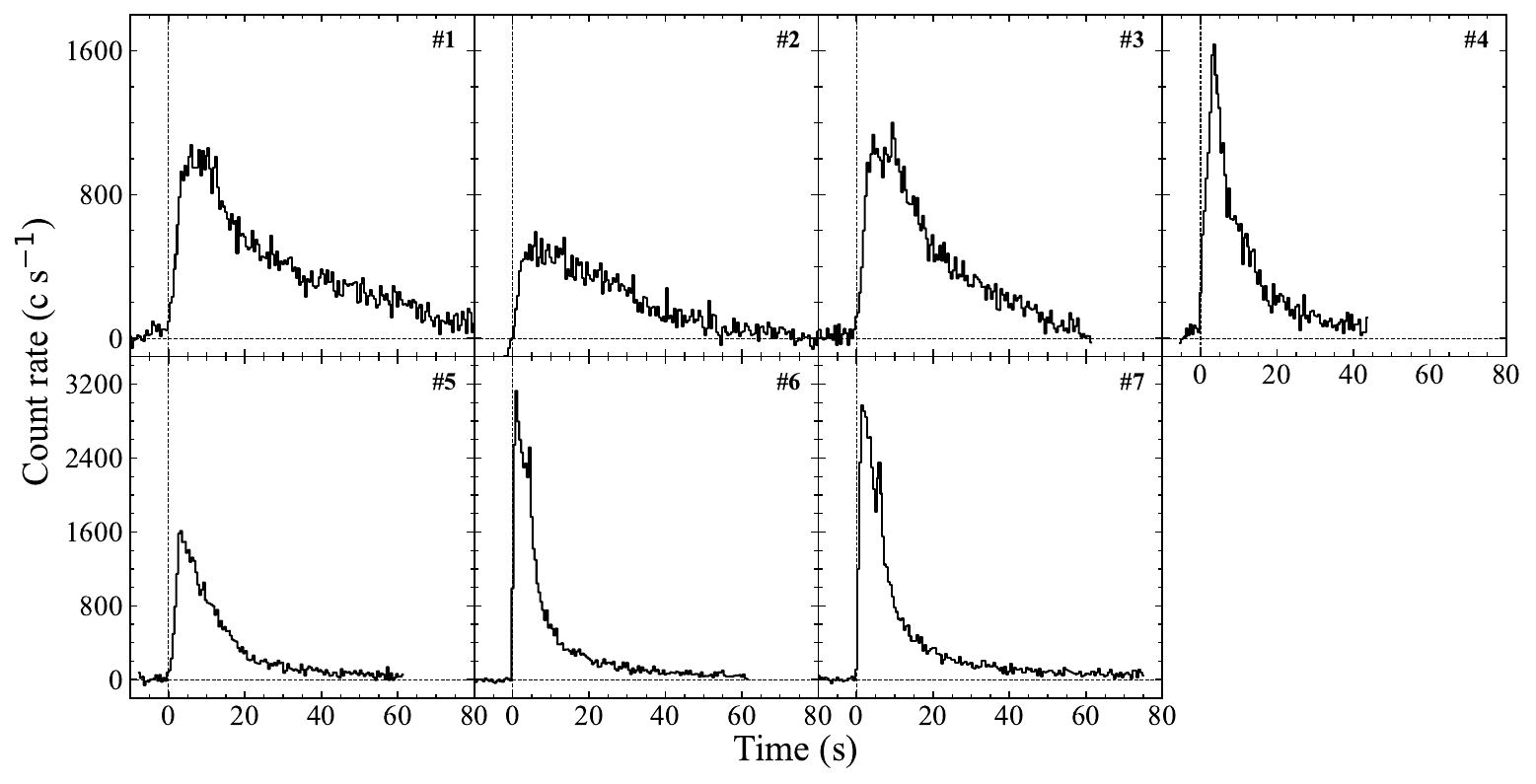}
s\caption{Light curves of the seven X-ray bursts from Terzan 5 X--3 observed with \nicer. Black lines are light curves in 0.5-10 keV  with time bin of 0.5 s. The vertical dotted line marks the onset time of each burst. The gray dashed dot represents the persistent emission, which were subtracted from the bursts.}
\label{fig:burst_lc}
\end{figure*}

The resulting net light curves, presented in Fig.~\ref{fig:burst_lc}, reveal a clear evolution in the burst properties over the course of the outburst. The first three bursts are morphologically similar, with relatively low peak count rates of
$<$1600\,counts\,s$^{-1}$. The subsequent two bursts (bursts~\#4 and \#5)
reach higher peak rates of $\sim$1800\,counts\,s$^{-1}$. The final two bursts
are distinct, exhibiting prominent double-peaked structures and significantly higher net count rates, peaking at $\sim$3000\,counts\,s$^{-1}$. A complete summary of these properties, along with the burst detected by \cxo\ during this outburst \citep{2023ATel15953....1H}, is provided in Table~\ref{table:burst_properties}.

We searched for coherent burst oscillations in the 0.5--10, 0.5--3, 3--6 keV event data for all seven bursts. Following standard techniques \citep[e.g.,][]{Bilous19,li2022ApJ}, we computed Leahy-normalized power spectra in the 50--2000\,Hz range using a 4-s sliding window FFT. No significant signal was detected. To place an upper limit on the oscillation amplitude, we calculated the fractional root-mean-square (rms) amplitude, $A_\mathrm{rms}$. Given the high count rates of the bursts, the contribution from the non-source background is negligible, and the formula simplifies to $A_\mathrm{rms} \approx\sqrt{P_\mathrm{s}/N_\mathrm{ph}}$, where $N_\mathrm{ph}$ is the total number of source photons in the interval and $P_\mathrm{s}$ is the signal power \citep{2019ApJ...878..145M, 2022AJ....163..130R}. The signal power was estimated from the measured power in the FFT, following the statistical methods of \citet{1975ApJS...29..285G} and \citet{1994ApJ...435..362V} \citep[see also][]{2019ApJS..245...19B,2022AJ....163..130R}. These methods are essential for correctly interpreting the power
spectrum of photon counting data. Specifically, \citet{1975ApJS...29..285G} first rigorously established that for a Leahy-normalized power spectrum, the power values arising from a pure Poisson process follow a chi-squared ($\chi^2$) distribution with 2 degrees of freedom. \citet{1975ApJS...29..285G} and \citet{1994ApJ...435..362V} provided a practical prescription for estimating the underlying signal power, $P_s$, and its statistical uncertainty from a measured power, $P_m$. From this analysis, we obtained an upper limit of 5.5\% on the fractional rms amplitude of any coherent signal.

\begin{table*}
\begin{center} 
\caption{Burst parameters' overview.  \label{table:burst_properties}}

\resizebox{\linewidth}{!}{\begin{tabular}{ccccccccccccccc} 
{\centering  Burst  } &
{\centering  Obs. Ids.$^{\mathrm{a}}$ } &
{\centering  Burst Onset$^{\mathrm{b}}$} &
{\centering  Peak Rate$^{\mathrm{c}}$} &
{\centering  $F_{\rm peak}^{\rm d}$  } &
{\centering  $E_{\rm b}$} &
{\centering  PRE  } &
{\centering  $F_{\rm TD}^{\rm e}$ } &
{\centering  $kT_{\rm TD}$ } &
{\centering  $\Delta T_{\mathrm{rec}}^{\rm f}$ } &
{\centering  $\Delta t_{\mathrm{rise}}^{\rm g}$ } &
{\centering  $\tau^{\rm h}$ } &
\\
 $\#$ & & (MJD) &($\mathrm{10^{3} ~ c ~ s^{-1}}$) &  &($10^{-7}~ \mathrm{erg~cm^{-2}}$)&  &  & (keV) &(hr) &(s)&(s) &\\ [0.01cm] \hline
1  &x01& 60004.28416 & 1.32 & $3.27\pm0.15$ & $4.56\pm{0.22}$ &N&       -       &       -       & -   &2.8& $13.94\pm{0.93}$ & \\
2  &x01 & 60004.34022 & 0.81 & $1.58\pm{0.08}$ & $3.25\pm{0.16}$ &N&       -       &       -       & 1.34     &2.8& $20.57\pm{1.45}$ &\\
3  &x01& 60004.92388 & 1.42 & $2.87\pm{0.12}$&$4.69\pm{0.27}$&N &-&-& 13.99 &2.9& $16.34\pm{1.16}$ &\\
4  &x06 & 60009.37851 & 2.41 & $5.02\pm{0.23}$&$3.07\pm{0.16}$&Y & $5.02\pm{0.23}$ & $3.25\pm{0.57}$ & 106.8  &2.4& $6.12\pm{0.42}$            &\\
-  &-   & 60021.64913 & -& - & -& -& -& -& 294.49& -& -&\\
5  &x15 & 60021.84144 & 1.98 & $3.92\pm{0.16}$ & $3.35\pm{0.13}$ & N &-&-& 4.61 &2.3& $8.55\pm{0.48}$            &\\
6  &x19 & 60025.57105 & 3.10 & $8.51\pm{0.39}$ & $3.96\pm{0.17}$ & Y &      $8.51\pm{0.39}$       &      $3.92\pm{0.83}$       & 89.51  &0.7& $4.56\pm{0.29}$ &\\
7  &x20 & 60026.97903 & 3.07 & $7.46\pm{0.33}$ & $4.91\pm{0.28}$ & Y &       $7.46\pm{0.33}$       &       $3.61\pm{1.12}$      & 33.59 &1.2& $6.58\pm{0.48}$            &\\
\hline  
\end{tabular} }

\end{center}

$^{\rm a}$ We only use the last two digits to represent \nicer\ Obs. Ids., so x=62036701. 

$^{\rm b}$ The start time of the Type I X-ray burst

$^{\rm c}$ The peak count rates were measured from the 0.5-second light curves in the energy range of 0.5–10 keV.

$^{\rm d}$ The bolometric peak flux of each burst is in units of $10^{-8}~ \mathrm{erg~s^{-1} cm^{-2}}$. 

$^{\rm e}$ The bolometric touchdown flux is in units of $10^{-8}~ \mathrm{erg~s^{-1} cm^{-2}}$.

$^{\rm f}$ The observed recurrence time.

$^{\rm g}$ The time of burst onset to its peak.

$^{\rm h}$ The decay time of bursts defined as the ratio of the burst fluence to its peak flux.

\end{table*}

\section{Spectral Analysis}   
\label{sec:spec_analysis}
We performed the spectra analysis using Xspec v12.12.0 \citep{Arnaud96}.  We generated the $\texttt{nibackgen3C50}$ background spectra \citep{Remillard21}, the ancillary response files ($\texttt{ARFs}$) and response matrix files ($\texttt{RMFs}$) using the tool $\texttt{nicerl3\text{-}spec}$. Since \nicer\ is an non-imaging instrument, the $\texttt{nibackgen3C50}$ model properly accounts for the non-X-ray particle background and the diffuse cosmic X-ray background, which contribute non-source counts. The errors of all parameters are quoted at the $1\sigma$ confidence level.

\subsection{Preburst persistent emission}
\label{Sec:spec_per}
For each burst, we extracted a 100 s pre-burst persistent emission spectrum. However, for the second burst, due to missing pre-burst observations, we used a 100 s post-burst persistent spectrum instead. We performed optimal binning for each persistent spectrum using {\tt ftgrouppha} as suggested by the \nicer\ team. The 3C50 background spectrum for each pre-burst was subtracted. Subsequently, we fitted the persistent emissions with a model consisting of an accretion disk multi-blackbody, \texttt{diskbb}, plus a blackbody, \texttt{bbodyrad}, modified by the Tübingen-Boulder absorption model, \texttt{TBabs}. Thus, the model is \texttt{TBabs}$\times$(\texttt{diskbb}+\texttt{bbodyrad}). 
The free parameters include the inner disk temperature, $T_{\rm in}$, and normalization, $K_{\rm diskbb}$, for ${\tt diskbb}$; the blackbody temperature, $kT_{\rm bb}$, and normalization, $K_{\rm bb}$, for ${\tt bbodyrad}$; as well as the equivalent hydrogen column density ($N_{\rm H}$) for ${\tt TBabs}$. The absorption column density were consistent across all spectra, $N_{\rm H}\sim 2.29\pm{0.03} \times 10^{22}\,\mathrm{cm}^{-2}$, which was also agreed with the result reported by \citet{2014ApJ...780..127B}. Therefore, in subsequent analysis, we fixed $N_{\mathrm{H}}$ to this value for all spectra during fitting. 
The model fitted the per-burst spectra well with $\chi^2$ per degree of freedom (dof), $\chi^2_{\nu} \approx 1$. The unabsorbed bolometric flux in the 0.1\textendash250 keV energy band were computed using the \texttt{cflux} tool. The best-fit parameters are listed in Table~\ref{table:preburst}.

\begin{table*}
\begin{center} 
\caption{ Best-fitted parameters of the pre-burst persistent spectra.  \label{table:preburst}}
{\begin{tabular}{ccccccccc} 
\hline %
{\centering \nicer} &
{\centering  $N_{\rm H}$  } &
{\centering  $T_{\rm in}$} &
{\centering  $K_{\rm diskbb}$} &
{\centering  $kT_{\rm bb}$} &
{\centering  $K_{\rm bb}$} &
{\centering  $\chi_\nu^{2}$(dof)}&
{\centering  $F_{\rm per}$$^{\mathrm{a}}$} &\\
(Obs. Id)& $(10^{22}~\mathrm{cm^{-2}})$ & $\mathrm{(keV)}$ &  & $\mathrm{(keV)}$ & &  & ($10^{-9}\enspace \mathrm{erg\enspace s^{-1}\enspace cm^{-2}}$) \\ [0.01cm] \hline
6203670101  &                 & $0.75\pm{0.05}$&$159_{-36}^{+45}$  & $1.79\pm{0.06}$  & $26\pm{3}$ & 1.02(107)&         $3.96\pm{0.03}$ \\
6203670101  &                 & $0.79\pm{0.07}$&$135_{-35}^{+44}$  & $1.74\pm{0.07}$  & $28\pm{4}$ & 1.01(107)&         $3.99\pm{0.03}$ \\
6203670101  &                 & $0.85\pm{0.07}$&$120_{-29}^{+36}$  & $1.82\pm{0.07}$  & $25\pm{4}$ & 1.16(109)&         $4.47\pm{0.03}$ \\
6203670106  & $2.29\pm{0.03}$         & $0.99\pm{0.07}$&$188_{-37}^{+43}$  & $1.72\pm{0.08}$  & $44\pm{9}$ & 1.11(118)&         $8.20\pm{0.03}$ \\
6203670115  &                 & $0.74\pm{0.03}$&$333_{-52}^{+61}$  & $1.50\pm{0.05}$  & $35\pm{6}$ & 0.86(115)&  $4.26\pm{0.03}$ \\
6203670119 &                 & $0.60\pm{0.02}$&$326_{-55}^{+67}$  & $1.43\pm{0.06}$  & $15\pm{2}$ & 0.96(104)&         $1.65\pm{0.03}$ \\
6203670120 &                 & $ 0.65\pm{0.03}$&$197_{-37}^{+46}$  & $2.32\pm{0.26}$  & $3\pm{0.8}$ & 0.84(101)&         $1.73\pm{0.03}$ \\
\hline

\hline
\end{tabular} }

\end{center}

$^{\mathrm{a}}$ The unabsorbed bolometric persistent flux in $0.1-250$ keV.

\end{table*}

\subsection{X-ray Burst Time-Resolved Spectroscopy}
\label{Sec:spec_burst}

We extracted the time-resolved spectra of the X-ray bursts, employing variable exposure time ranging from 0.125 to 4 s to ensure a minimum of 1000 counts including persistent emission for each spectrum. The burst spectra were grouped with a minimum of 20 counts per channel using the \texttt{grappha} tool. The hydrogen column density was fixed at $2.29\times10^{22}$ cm$^{-2}$ from the persistent spectral fitting, see Sect.~\ref{Sec:spec_per}. For each burst, the 3C50 background spectrum was subtracted.

We performed time-resolved spectral analysis of all seven X-ray bursts. To account for both the burst emission and any potential enhancement of the accretion flow via the Poynting-Robertson drag, we fitted each time-resolved spectrum with the model \texttt{TBabs $\times$ (bbodyrad +  $f_a\times$(diskbb+bbodyrad))}. Here, \texttt{bbodyrad}
represents the burst emission. The component (\texttt{diskbb} + \texttt{bbodyrad}) accounts for the pre-burst persistent emission, with the parameters fixed to the best-fit values listed in Table \ref{table:preburst}. 
We assumed that only the amplitude of the persistent emission during the burst increases by a scaling factor $f_a$ modeled by \texttt{constant} in Xspec, while the shape of persistent spectra did not change. We used an F-test to justify the inclusion of the enhancement factor $f_a$ as a free parameter in each time-resolved burst spectrum. The parameter was allowed to vary only if the F-test indicated a significant improvement ($p < 0.003$, corresponding to significant level of $>3\sigma$) over a model with $f_a$ frozen at 1. Otherwise, the simpler model with $f_a=1$ was adopted.

\begin{figure*}
\centering
\includegraphics[width=18cm]{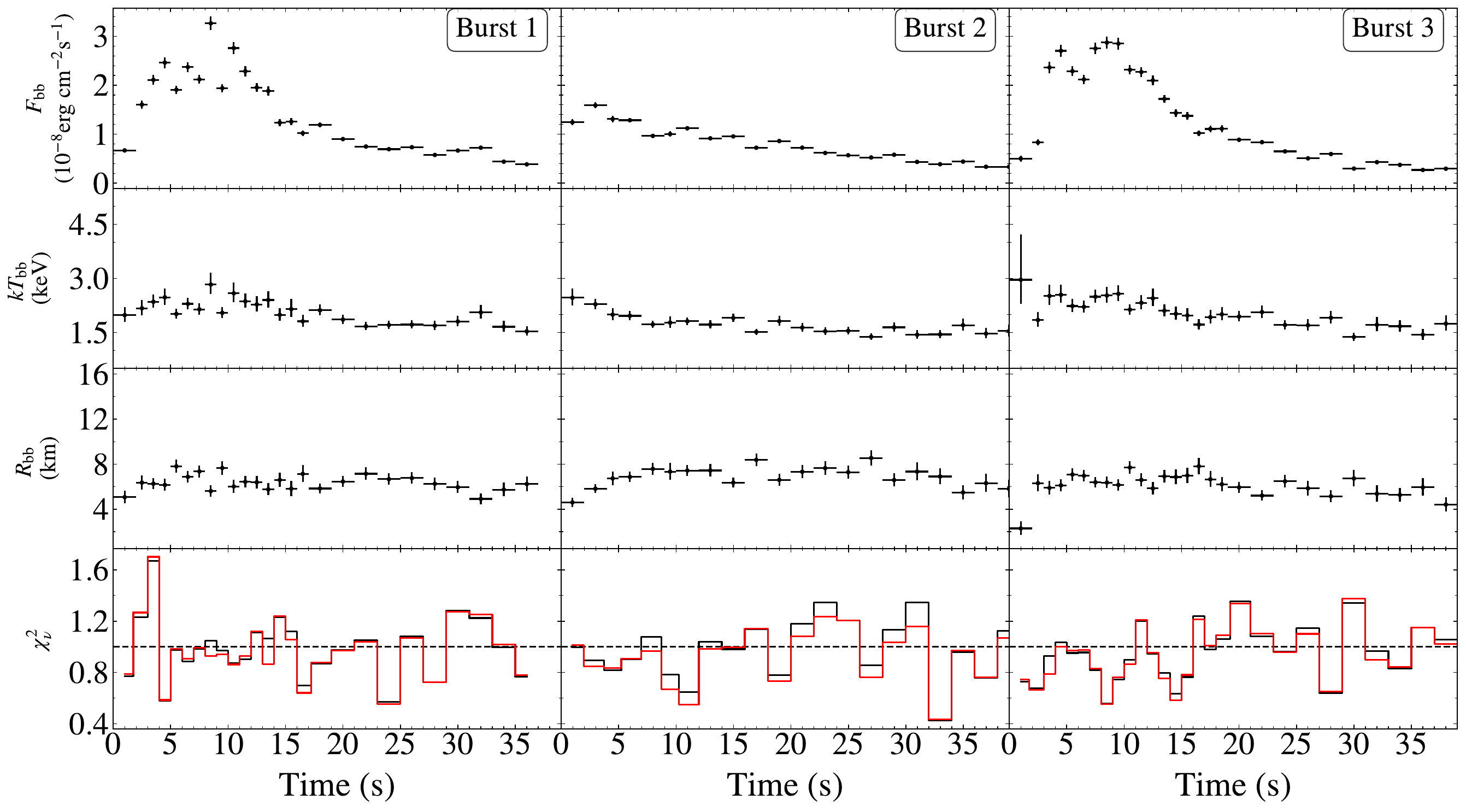}
\caption{The best-fitted parameters of the spectra from bursts \#1-3 using the model \texttt{TBabs} $\times$ (\texttt{bbodyrad} + $f_a$ $\times$ (\texttt{diskbb} + \texttt{bbodyrad})). From top to bottom, we show the bolometric blackbody flux, $F_{\text{bb}}$, blackbody temperature, $kT_{\text{bb}}$, the blackbody radius, $R_{\rm bb}$ and the goodness of fit, $\chi_\nu^{2}$. In the bottom panel, the red and black lines indicate the $\chi_\nu^{2}$ with $f_a$ free and fixed at 1, respectively. Since allowing $f_a$ to vary did not improve the fit, it was fixed at 1.}
\label{fig:fa1_3}
\end{figure*}

\begin{figure}
\centering
\includegraphics[width=0.5\textwidth]{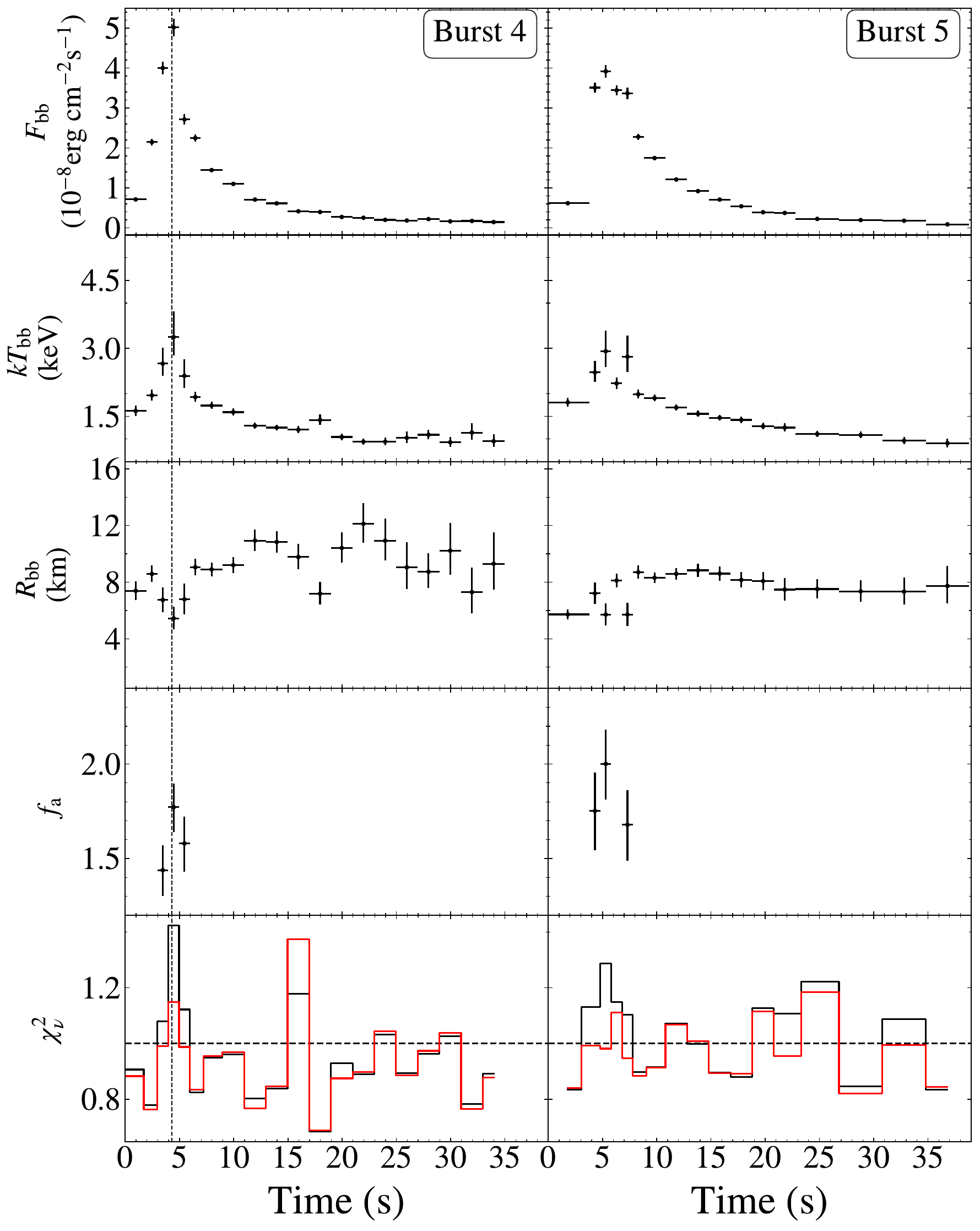}
\caption{Same as Fig.~\ref{fig:fa1_3}, but for bursts \#4 (left) and \#5 (right). In the forth panel, we show the factor $f_a$, which was set as free parameter improving the fit. The vertical dashed line indicates the touchdown moment for the PRE candidate, burst \#4.}
\label{fig:fa4_5}
\end{figure}

\begin{figure}
\centering
\includegraphics[width=0.5\textwidth]{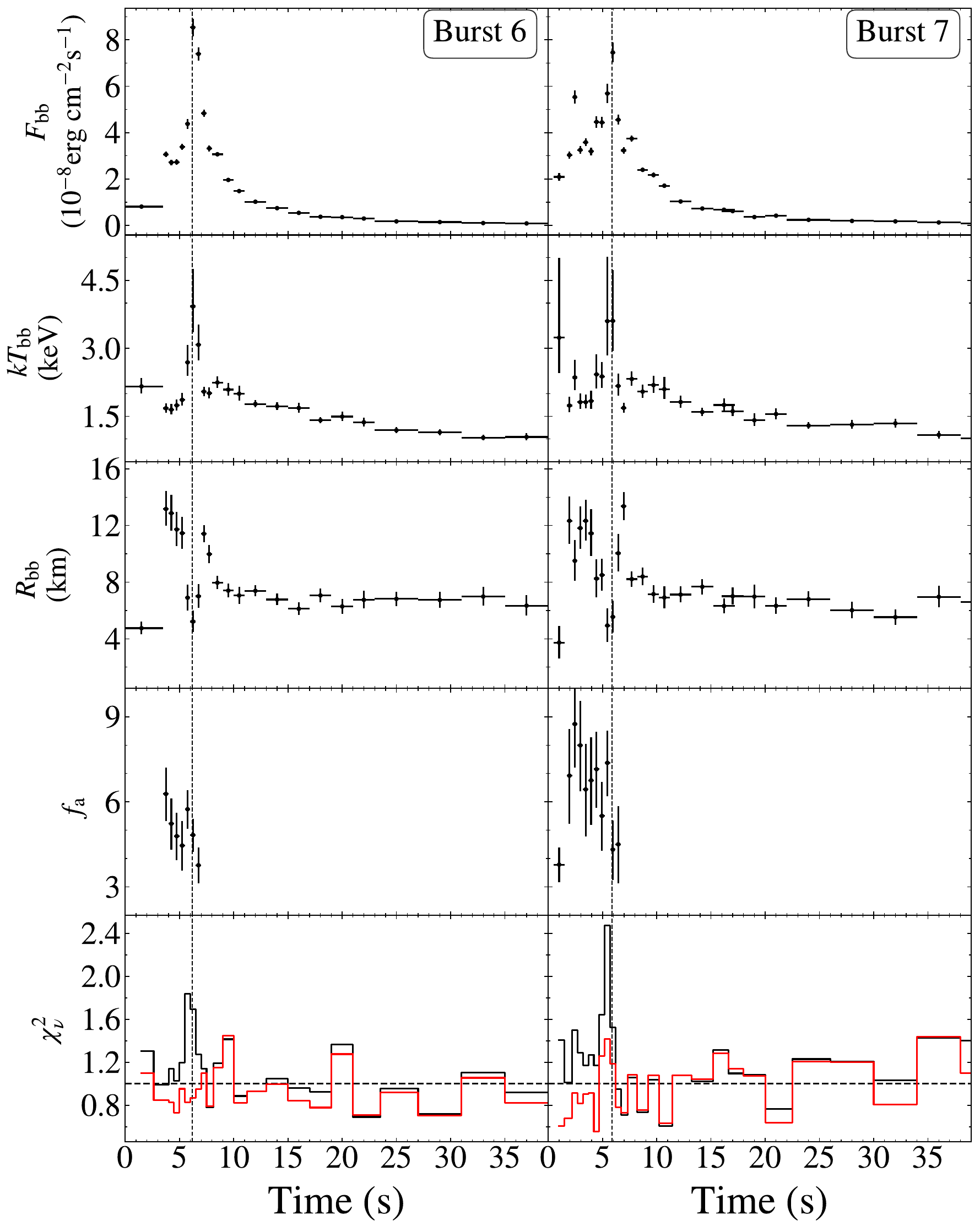}
\caption{Same as Fig.~\ref{fig:fa1_3}, but for bursts \#6 (left) and \#7 (right). The vertical dashed line indicates the touchdown moment.}
    \label{fig:fa6_7}
\end{figure}

Most of burst spectra are well-described by the $f_a$ model
with $\chi_\nu^{2}$ $\sim1$, see Figs.~\ref{fig:fa1_3}, \ref{fig:fa4_5} and \ref{fig:fa6_7}. For the first two bursts, the $f_a$ parameter is consistent with unity throughout their evolution. This indicates that the persistent emission was not significantly enhanced during these bursts. For bursts \#3, \#4, and \#5, we detected a modest and transient enhancement of the persistent flux. As shown in the fourth panels in Fig.~\ref{fig:fa4_5}, the $f_a$ factor was approximately $1.5-2$ only around the peak of these bursts.  For bursts \#6 and \#7, the enhancement factor $f_a$ was more significant, reaching  $6-8$ also around the burst peak.

For bursts \#1-3 and \#5, the bolometric flux raised to a peak of approximately $1.6-3.9 \times 10^{-8}$\,erg\,cm$^{-2}$\,s$^{-1}$ as the blackbody temperature reached a maximum of $\sim2-3$ keV. Subsequently, the bursts entered a cooling phase where the temperature steadily decreased over $20-30$ s. Crucially, the apparent blackbody radius remained largely constant at a value of $4-6$ km during the early phase, suggesting the absence of PRE and classified as non-PRE bursts. For burst \#4, the blackbody radius reached to $\sim$8 km, then dropped to $\sim$5 km, meanwhile, the bolometric flux and the blackbody temperature were peaked at $5.02 \times 10^{-8}$\,erg\,cm$^{-2}$\,s$^{-1}$ and 3.25 keV, respectively. We regarded it as a PRE burst candidate, even though the blackbody radius did not expanded significantly.

The best-fitted parameters of last two bursts are very similar, see Fig.~\ref{fig:fa6_7}. The apparent blackbody radius, $R_\mathrm{bb}$, underwent a rapid expansion in $\sim1$ s, increasing to over 13 km during the initial phase while the temperature, $kT_\mathrm{bb}$, showed a corresponding drop. Then the radius decreased to $\sim 4$ km, meanwhile, the temperature increased to a peak of $\sim 4$ keV, indicating the NS photosphere was contracting. At this moment, the bolometric flux raised to a peak at $\sim8 \times 10^{-8}$\,erg\,cm$^{-2}$\,s$^{-1}$. Afterwards, the photosphere was cooling and the bolometric flux decayed exponentially. These signatures can identify these bursts as PRE.

Following the time-resolved spectral analysis, we derived several properties for each burst, which are presented in Table~\ref{table:burst_properties}.
The burst fluence, $E_\mathrm{b}$, was calculated by numerical integrating the blackbody flux over the entire burst duration. We computed the decay timescale, $\tau$, as the ratio of the burst fluence to the peak flux, $\tau = E_\mathrm{b} / F_\mathrm{peak}$.  Finally, the observed recurrence time, $\Delta T_\mathrm{obs}$, between consecutive bursts was calculated simply as the difference between their onset time.

\section{Discussion}
\label{Sec:discussion}
We analyzed the observations of Terzan 5 X--3 in the 2023 outburst, constructing light curves and HID based on the \nicer\ and \maxi\ data. The outburst behavior is remarkably similar to its previous outburst in 2012 \citep{2014ApJ...780..127B}. We found seven type I X-ray bursts and performed time-resolved burst spectral analysis. The persistent spectrum prior to each burst, in the energy range of 1--10 keV, can be well fitted with a \texttt{Tbabs $\times$(diskbb + bbodyrad)} model. 
From the time-resolved spectral studies with the $f_a$ model, we found that the strength of the persistent emission enhancement, quantified by $f_a$, scales directly with the burst luminosity.
The two faintest non-PRE bursts showed no significant enhancement ($f_a \approx 1$);
the two more luminous non-PRE bursts and one PRE candidate burst exhibited a modest and transient enhancement ($f_a \approx 1.5-2$); and the two most powerful PRE bursts required a strong
enhancement ($f_a \approx 6-8$) during the burst peak. This clear trend
provides compelling, quantitative evidence for a physical scenario where the
strength of the burst-disk interaction, likely caused by the Poynting-Robertson drag, is a direct function of the thermonuclear flash's radiative power \citep{Worpel13,Worpel15}.

We found systematic difference in the apparent blackbody radii, $R_\mathrm{bb}$, during the cooling tails of these bursts (see Figs.~\ref{fig:fa1_3}, \ref{fig:fa4_5} and \ref{fig:fa6_7}). The bursts \#1-3 and \#6-7 exhibited a consistent radius of approximately $R_\mathrm{bb} \approx 6$\,km during the cooling tail. In contrast, bursts \#4 and \#5 showed a systematically larger radius during the cooling tail, $R_\mathrm{bb} \approx 8-10$\,km. A plausible origin for this behavior is a change in the spectral color-correction factor, $f_\mathrm{c}$. The apparent blackbody radius, $R_\mathrm{bb}$, is related to the true NS radius, $R_\mathrm{NS}$, by $R_\mathrm{NS} = f_\mathrm{c}^2 R_\mathrm{bb}$, if the emission is from the whole NS surface. The color-correction factor accounts for the fact that the emergent spectrum from a
NS atmosphere is not a perfect blackbody. Theoretical models show that
$f_\mathrm{c}$ depends on the atmospheric composition, temperature, burst flux, and surface gravity, and can also be affected by the accretion environment \citep[e.g.,][]{Suleimanov11, Suleimanov12, Poutanen14}. Bursts \#4 and \#5 occurred at a significantly higher persistent luminosity and in a spectrally softer state than the first three bursts (see Fig.~\ref{Fig:HID}). It is possible that the different accretion environment and higher surface temperature altered the atmospheric structure, leading to a different color-correction factor (i.e., a slightly smaller $f_\mathrm{c}$) for these bursts. This would manifest as a larger apparent radius, $R_\mathrm{bb}$, as we observe.

\subsection{Two Burst Types: Evidence from Morphology}
During its 2012 outburst, Terzan 5 X--3 exhibited a single non-PRE burst with a rise time of 3 s and a long decay of $\tau \approx 16-29$ s. This duration is a clear signature of mixture of hydrogen and helium burning, which confirmed the binary is hosting a NS accreting from a hydrogen-rich companion \citep{2014ApJ...780..127B}. 

In the 2023 outburst, the properties of the seven X-ray bursts strongly suggest a change in the burning fuel. The first five bursts were non-PRE events with rise times of 2--3\,s and decay timescales ($\tau = E_\mathrm{b}/F_\mathrm{peak}$) of 6--21\,s (see Table~\ref{table:burst_properties}).  Such long durations are characteristic of mixed hydrogen/helium burning, where the energy release is moderated by $\beta$-limited CNO cycle reactions.  In stark contrast, the final two bursts of the 2023 outburst were powerful PRE events. They exhibited short rise times of $\sim$1\,s and rapid decays ($\tau \approx 5$\,s). This impulsive morphology is the classic signature of unstable pure helium ignition via the $3\alpha$ process, which proceeds on a much faster reaction timescale. Based on this strong morphological evidence, we suggest that the fuel for the bursts transitioned from a mixed hydrogen/helium to a pure He burning over the course of the outburst.

\subsection{The PRE Bursts: A Standard Candle Test for Fuel Composition}
\label{sec:pre_test}

Our fuel composition hypothesis can be tested using the PRE bursts. The bolometric flux at the touchdown moment of a PRE burst is expected to correspond to the Eddington luminosity ($L_\mathrm{Edd}$), making these events as standard candles \citep{Kuulkers03}. Since $L_\mathrm{Edd}$ depends on the atmospheric hydrogen fraction, $X$, we can test our hypothesis by comparing the derived distance for different fuel compositions to the known distance of Terzan~5. The observed average touchdown flux of the two PRE bursts (\#6--7) is
$F_{\rm TD} = (8.0 \pm 0.3) \times 10^{-8}~\mathrm{erg~s^{-1}~cm^{-2}}$.
At the precise distance measurement of $6.62 \pm 0.15$\,kpc to Terzan~5 \citep{2022ApJ...941...22M}, the luminosity at the touchdown is 
$L_{\rm TD}=4\pi D^2F_{\rm TD} = (4.2 \pm 0.3) \times 10^{38}~\mathrm{erg~s^{-1}}$. 
We compare this value to the theoretical Eddington luminosity, which includes a temperature-dependent correction for the opacity of the scattering atmosphere \citep{Lewin93, Suleimanov12, Poutanen17}, 

\begin{equation}
\begin{split}
L_{\rm Edd,~\infty }
&=\frac{8\pi Gm_{p} M_{\rm NS}c[1+(kT/39.4~\rm{keV})^{0.976}]}{\sigma_{\rm T}(1+X)(1+z)}  \\
&=2.7\times 10^{38}\Bigr (\frac{M_{\rm NS}}{1.4M_\odot }\Bigr )\frac{1+(kT/39.4~{\rm keV})^{0.976} }{(1+X)}\\
&\quad \times \Bigl (\frac{1+z}{1.31}\Bigr)^{-1}~\mathrm{erg~s^{-1}}, 
\end{split}
\label{eq:L_edd} 
\end{equation}
where $m_{\rm p}$ is the mass of the proton, $\sigma_{\rm T}$ is the Thompson scattering cross section, $kT$ is the effective temperature of the atmosphere in the unit of keV, $G$ is the gravitational constant, the gravitational redshift factor $1+z=(1-2GM_{\rm NS}/Rc^2)^{-1/2}$, and $X$ is the mass fraction of hydrogen. We assume the blackbody temperature at the touchdown moment as the effective temperature, this is, $kT=kT_{\rm TD}$. For our observed touchdown temperature of $kT \approx 4.0$\,keV, this
equation simplifies to $L_{\rm Edd,~\infty} \approx 3.0 \times 10^{38}
(M_{\rm NS}/1.4\,M_\odot)(1.31/1+z)/(1+X)$\,erg\,s$^{-1}$.  To produce the observed
luminosity would require a NS mass of $M_{\rm NS} \approx 3.3\,M_\odot$ for $X=0.7$ and  $M_{\rm NS} \approx 2\,M_\odot$ for $X=0$, assuming the NS radius of 10--13 km \citep[e.g.,][]{Dittmann24,Vinciguerra24}. 
Such a high mass in the former case significantly exceeds the maximum known mass for a NS, allowing us to confidently reject this scenario \citep[e.g.,][]{Kiziltan13}.
For the PRE candidate, burst \#4, its observed peak flux of
$F_{\rm TD} \approx 5 \times 10^{-8}$\,erg\,s$^{-1}$\,cm$^{-2}$ corresponds to an observed luminosity of $L_{\rm TD} \approx 2.76 \times 10^{38}$\,erg\,s$^{-1}$. To produce this lower observed luminosity in mixed hydrogen/helium with $X=0.7$, it also requires a NS mass of $M_{\rm NS} \approx 2\,M_\odot$, consistent with the PRE bursts in pure He environment. Therefore, our analysis strongly
indicates that the observed PRE events reflect a transition from a mixed-fuel PRE burst to more powerful pure-helium PRE bursts, all occurring on a massive NS.

\subsection{The Non-PRE Bursts: A Recurrence Time Consistency Check}
\label{sec:nonpre_check}

We perform a consistency check on the non-PRE bursts using their recurrence
times. From the pre-burst persistent flux, the local mass accretion rate can be calculated via the relation \citep{Galloway08}, 
\begin{equation}
\begin{split}
    \Dot{m}
    &=\frac{L_\mathrm{ {per}}(1+z)}{ 4\pi R_{\rm NS}^{2}(GM_{\rm NS}/R_{\rm NS})}\\
    &\approx 2.94\times 10^{3}\biggl(\frac{F_\mathrm{{per}}}{10^{-9}\mathrm{~ergs ~ cm^{-2}~s^{-1}}}\biggr)\biggl(\frac{d}{6.62\mathrm{~ kpc}}\biggr)^{2}\biggl(\frac{M_{\rm NS}}{1.4M_{\odot}}\biggr)^{-1}\\
    &\quad\times\biggl(\frac{1+z}{1.31} \biggr)\biggl(\frac{R\mathrm{_{NS}}}{10\mathrm{~ km}}\biggr)^{-1}\mathrm{g~cm^{-2}}\mathrm{~s^{-1}},
    \label{eq:lo_accration}
\end{split}      
\end{equation}
where $F_{\rm{per}}$ is the persistent flux, $d$ is the source distance, and the NS mass  $M_{\rm{NS}}=1.4M_\odot$ and radius $R_{\rm{NS}}=10$ km. 
The results are listed in Table~\ref{table:Calculated}. Please note that $\dot{m}$ can vary by $\sim$30\% over a plausible range
of NS parameters ($M_{\rm NS} = 1.4-2.0\,M_\odot$, $R_{\rm NS} = 10-13$\,km). Nevertheless, this uncertainty does not alter  the main conclusions presented below.
The predicted recurrence time, $\Delta t_{\rm rec}$, depends on the ignition
depth, $y_{\rm ign}$, and the local mass accretion rate, $\dot{m}$,
\begin{equation}
y_{\rm ign}=\frac{4\pi E_\mathrm{{b}}d^{2}(1+z)}{4\pi R_{\rm NS}^{2}Q\mathrm{_{nuc}}}, \label{eq:ign}
\end{equation} 
and
\begin{equation}
    \Delta t_{\rm rec} = \frac{y_{\rm ign}}{\dot{m}}(1+z),
\end{equation}
where the nuclear energy generated is $Q_\mathrm{{nuc}} \approx 1.31+6.95X -1.92X^{2}\mathrm{~MeV~nucleon^{-1}}\approx 4.98 \mathrm{~MeV~nucleon^{-1}}$ for solar composition ($X=0.7$)   and $Q_\mathrm{{nuc}} \approx 1.31 \mathrm{~MeV~nucleon^{-1}}$ for helium ($X=0$)  \citep{Goodwin19}. We used $d=6.62\pm 0.15$ kpc to determine  $y_\mathrm{ign}$ and $ \Delta t_{\rm rec}$.

For burst~\#2, the observed recurrence time since burst~\#1 was 1.34\,hr. Assuming a mixed-fuel composition ($X=0.7$), the observed fluence and persistent flux predict a recurrence time of $1.03\pm0.26$ hr. This is in good agreement with the observed value, strongly supporting our conclusion that the non-PRE bursts are fueled by a hydrogen/helium mixture. We also computed the ratio of the 
persistent flux to the burst fluence, $\alpha=T_{\rm rec}F_{\rm per}/E_{\rm b}$, which is  $66\pm4$ for  burst \#2, also consistent with the mixed hydrogen/helium burst  \citep{Li21,Galloway08}.

\begin{figure}
\centering
\includegraphics[width=9cm]{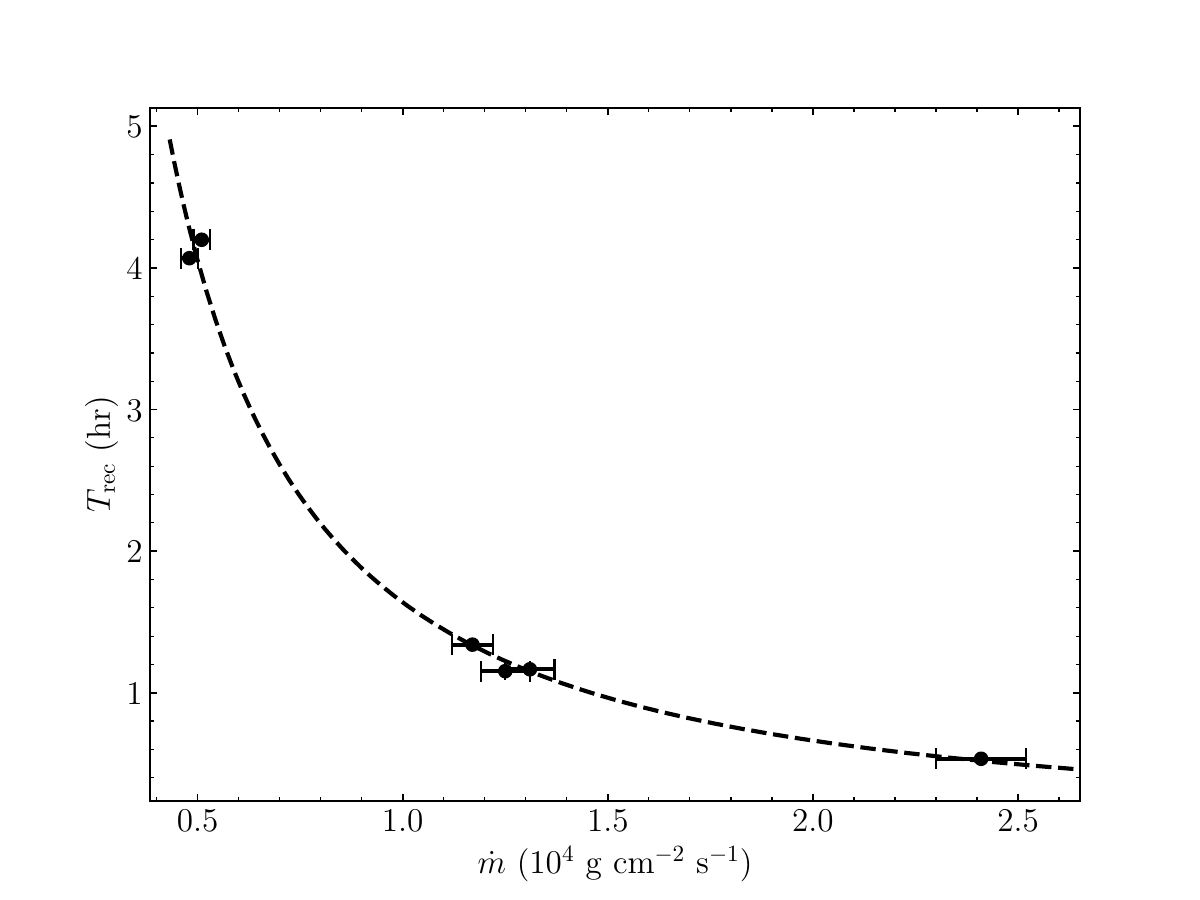}
\caption{Recurrence time versus local mass accretion rate. The dashed line represent the best-fit power-law model for all bursts $\Delta T_\mathrm{rec} \propto \dot{m}^{-1.31 \pm 0.04}$.}
\label{fig:mdot-trec}
\end{figure}

In addition, we can examine the relation between the recurrence time ($\Delta T_\mathrm{rec}$) and the local mass accretion rate ($\dot{m}$) across the entire outburst. Due to observational gaps, the measured recurrence time between bursts, $\Delta T_\mathrm{rec}$, is an upper limit on the true value. If $N$ bursts were missed between two observed successive bursts, we corrected the observed recurrence time via $\Delta T_\mathrm{rec} / (N+1)$.
Specifically, we chose $N=0, 11, 199, 3, 21, 7$ for the six intervals
between our seven observed bursts, with the largest number of missed bursts corresponding to the interval of burst \#4.
A power-law fit to these data reveals a strong anti-correlation,
$\Delta T_\mathrm{rec} \propto \dot{m}^{-1.31 \pm 0.04}$, see Fig.~\ref{fig:mdot-trec}.

This single power-law fit effectively describes the overall trend across the outburst. However, it should be noted that this is a simplification. Theoretical models and prior observations indicate that the power-law index itself depends on the burning regime, with an expected index of approximately $-1$ for mixed hydrogen/helium bursts, steepening to  $\sim-3$ for pure helium bursts \citep{CB2000,2011MNRAS.410..179C, Linares12,Li18b}. Our power-law index of $-1.31$ can be interpreted as a sample-averaged value across both regimes. While our dataset, which includes both burst types, is too small to constrain the indices for the different fuel compositions independently, the strong overall anti-correlation provides global support for our interpretation of the burst behavior.

\subsection{A burning transition from mixed fuel to pure helium}

Our analysis, combining burst morphology, distance constraints from PRE touchdown fluxes, and recurrence time calculations, paints a remarkably self-consistent picture of a fuel transition during the 2023 outburst of Terzan 5 X--3. The source began by producing mixed hydrogen/helium bursts from a hydrogen-rich accreted layer at accretion rates of  $0.13-0.27\dot{m}_{\rm Edd}$. As the outburst decayed and the accretion rate dropped to $\sim0.03\dot{m}_{\rm Edd}$, the burning regime transitioned. This observed behavior aligns with theoretical predictions, which place the transition threshold between these two burning regimes at $\sim 0.1\dot{m}_\mathrm{Edd}$ \citep{Strohmayer06}.

Observing a clear transition from a mixed-fuel to a pure-helium burning regime within a single outburst from NS LMXB is a rare phenomenon.  In ultracompact X-ray binaries (UCXB) with helium-rich donors, such as 4U\,1820--30, the accreted material is hydrogen-deficient by definition, so pure helium bursts are observed \citep{2024A&A...683A..93Y} (in rare case, it showed superburst powered by unstable carbon burning; \citealt{2025ApJ...982...18P}). However, bursting activity is not guaranteed. For instance, the UCXB Swift J1756.9--2508 has undergone two well-monitored outbursts, but both were relatively faint and no thermonuclear bursts were detected at all \citep{Li21}. While some of these systems at very low accretion rates ($\sim 0.01\dot{m}_{\rm Edd}$) can produce intermediate-duration bursts from the ignition of a thick helium layer \citep{2023A&A...670A..87L}. In these sources, a transition from mixed hydrogen/helium to helium burning is unlikely.

As systems where the mass accretion rate evolves significantly, transient LMXBs with hydrogen-rich donors are the prime candidates for observing burning transition. Such burning transition, driven by a decreasing accretion rate, has been explored in detail in a few transient LMXBs. \citet{2011MNRAS.410..179C}  studied a 6-month outburst of IGR J17473--2721 and identified a clear hysteresis in burst activity; mixed-fuel bursts dominated the rise, but after a month-long pause at the outburst peak, the bursting resumed with shorter, more energetic flashes consistent with pure helium burning at a lower persistent flux. Similarly, \citet{Linares12} analyzed IGR J17480--2446 (also in Terzan 5) and found direct evidence for a transition between pure helium and mixed hydrogen/helium ignition regimes at a persistent luminosity of approximately 30\% of the Eddington limit.

The monitoring of the Terzan 5 X--3 outburst by \nicer\ was therefore crucial, even though the \nicer\ observations only covered a low observational
duty cycle of $\sim5\%$ during the outburst. It provided the rare opportunity to witness the thermonuclear burning evolution, offering a valuable chance for studying the interplay between accretion rate and the conditions that trigger different thermonuclear bursts on NS surface.

\section{Summary}
\label{Sec:conclusion}
We have presented a detailed analysis of seven thermonuclear X-ray bursts from Terzan~5~X-3 observed by \nicer\ during its 2023 outburst. By performing time-resolved spectral analysis for all seven bursts, we have uncovered a clear physical evolution in the thermonuclear burning process and the burst-disk interaction. We summary the results as follows:

\begin{enumerate}
    \item We classify the seven bursts into two distinct types: four standard non-photospheric radius expansion (non-PRE) bursts and one PRE-candidate burst (burst~\#4) occurred in a mixed hydrogen/helium environment; and two powerful PRE bursts powered by pure helium (bursts \#6 and \#7).

    \item The strength of the interaction between the burst and the accretion disk, modeled with a persistent emission enhancement factor ($f_a$), scales directly with the burst peak luminosity. The enhancement is absent ($f_a \approx 1$) in the faintest bursts, becomes modest ($f_a \approx 1.5-2$) for the more luminous non-PRE bursts and the mixed-fuel PRE candidate, and is very strong ($f_a \approx 6-8$) during the pure helium PRE events. This provides quantitative evidence that the Poynting-Robertson drag effect is a direct function of the burst's radiative power.

    \item We confirm a transition in the burning regime from mixed hydrogen/helium fuel to pure helium, which occurs as the persistent accretion rate drops below $\sim$0.1\,$\dot{m}_\mathrm{Edd}$, in agreement with theoretical predictions.

    \item Our fuel composition and burst classifications are validated by a self-consistency check. The observed peak luminosities of both the mixed-fuel PRE candidate and the pure helium PRE bursts are consistent with reaching their respective, composition-dependent Eddington limits on the same massive NS of $\sim$2\,$M_\odot$.
\end{enumerate}

\begin{acknowledgements}
This work was supported by the Major Science and Technology Program of Xinjiang Uygur Autonomous Region (No. 2022A03013-3), and the science and technology innovation Program of Hunan Province (No.2024JC0001). Z.L. and Y.Y.P were supported by National Natural Science Foundation of China (12130342, 12273030).  This research has made use of data obtained from the High Energy Astrophysics Science Archive Research Center (HEASARC), provided by NASA’s Goddard Space Flight Center.
\end{acknowledgements}

\begin{table*}
\begin{center} 
\caption{Calculated parameters.  \label{table:Calculated}}
\resizebox{\linewidth}{!}{\begin{tabular}{ccccccccccccc} 
 & &  &$X=0.$ & && & $X=0.7$ && & \\
\cline{3-5} \cline{7-9}\\
{\centering  Burst } &
{\centering  $\Dot{m}$} &
{\centering  $\Dot{m}/\Dot{m}_{\rm Edd}$} &
{\centering  $y_{\rm ign}$} &
{\centering  $\Delta t_\mathrm{{rec}}^{a}$ } & 
&
{\centering  $\Dot{m}/\Dot{m}_{\rm Edd}$} &
{\centering  $y_{\rm ign}$} &
{\centering  $\Delta t_\mathrm{{rec}}^{a}$ } & \\
 $\#$ & ($\mathrm{10^{4} ~ g ~ cm^{-2} ~ s^{-1}}$) &\% & ($10^{8}\mathrm{~g~ cm^{-2}}$) &(hr)&&\%&($10^{8}\mathrm{~g~ cm^{-2}}$) &(hr) &\\ [0.01cm] \hline
1&$1.16\pm{0.05}$&$7.75\pm{0.35}$&$1.97\pm{0.13}$&$6.18\pm{0.50}$& &$13.18\pm{0.60}$&$0.52\pm{0.03}$&$1.63\pm{0.12}$&\\
2&$1.17\pm{0.05}$&$7.82\pm{0.36}$&$1.41\pm{0.09}$&$4.39\pm{0.35}$& &$13.30\pm{0.61}$&$0.37\pm{0.02}$&$1.15\pm{0.08}$&\\
3&$1.31\pm{0.06}$&$8.76\pm{0.40}$&$2.03\pm{0.15}$&$5.64\pm{0.49}$& &$14.89\pm{0.68}$&$0.53\pm{0.04}$&$1.47\pm{0.13}$&\\
4&$2.41\pm{0.11}$&$16.11\pm{0.73}$&$1.33\pm{0.09}$&$2.01\pm{0.16}$& &$27.39\pm{1.24}$&$0.35\pm{0.02}$&$0.53\pm{0.04}$&\\
5&$1.25\pm{0.06}$&$8.36\pm{0.38}$&$1.45\pm{0.09}$&$4.22\pm{0.33}$& &$14.20\pm{0.65}$&$0.38\pm{0.02}$&$1.11\pm{0.08}$&\\
6&$0.48\pm{0.02}$&$3.24\pm{0.17}$&$1.71\pm{0.11}$&$12.96\pm{1.06}$& &$5.50\pm{0.27}$&$0.45\pm{0.03}$&$3.41\pm{0.28}$&\\
7&$0.51\pm{0.02}$&$3.40\pm{0.17}$&$2.12\pm{0.15}$&$15.13\pm{1.30}$& &$5.77\pm{0.28}$&$0.56\pm{0.04}$&$4.00\pm{0.35}$&\\
\hline  
\end{tabular}} 
\end{center}
$^{a}~\Delta t_\mathrm{{rec}}$ is the estimate for the recurrence time (see Sec.~\ref{sec:nonpre_check} for more details)\\
\end{table*}

\bibliography{name}{}
\bibliographystyle{aa}

\end{document}